# Brain's Statistical Learning and its Embodied Mechanisms Towards Understanding Creativity and Individuality in Music

Tatsuya Daikoku

Graduate School of Information Science and Technology, The University of Tokyo, Tokyo, Japan
daikoku.tatsuya@mail.u-tokyo.ac.jp

# Abstract

Music is a universal feature of human culture, linked to embodied cognitive functions that drive learning, action, and the emergence of creativity and individuality. Evidence highlights the critical role of statistical learning—an implicit cognitive process of the brain—in musical creativity and individuality. Despite its significance, the precise neural and computational mechanisms underpinning these dynamic and embodied cognitive processes remain poorly understood. This paper discusses how individuality and creativity emerge within the framework of the brain's statistical learning, drawing on a series of neural and computational studies. This work offers perspectives on the mechanisms driving the heterogeneous nature of statistical learning abilities and embodied mechanisms and provides a framework to explain the paradoxical phenomenon where individuals with specific cognitive traits that limit certain perceptual abilities excel in creative domains.

# 1 Introduction

Music has been a fundamental aspect of human culture, deeply embedded in the embodied cognitive and emotional frameworks that define our species [1-3]. The connection between music, the brain, and the body extends beyond mere appreciation; it plays a critical role in learning, action, and the emergence of creativity and individuality [4].

Recent studies have suggested that the brain's implicit learning mechanism, called as "*statistical learning*" plays an important role in shaping musical creativity and individuality [5-7] and embodied cognition underlying musical emotion [8]. Statistical learning [9], which involves detecting patterns and regularities in sensory input, emerges as an innate capacity from early brain development [10] and continues to shape individual differences throughout life [11-13]. This adaptive process constructs unique frameworks based on the diversity of sensory experiences, effectively tailoring an individual's embodied cognitive capabilities. Such individuality influences not only how people engage with music but also how creativity manifests in their musical expressions [6, 14]. Despite the vital importance of statistical learning, its exact neural and computational mechanisms remain unclear, making it difficult to fully understand how it supports creativity and shapes individuality.

This paper discusses how individuality and creativity emerge within the framework of statistical learning of the brain, drawing on a series of neural and computational studies. Building on these interdisciplinary insights, this paper introduces a computational framework called as Hierarchical Bayesian Statistical Learning (HBSL) model [15], mimicking the brain's hierarchical processes of statistical learning. This provides a novel perspective on the diverse nature of statistical learning abilities and embodied cognitive mechanisms, shedding light on understanding how creativity and individuality could emerge through statistical learning processes.

# 2 Predictive Processing based on Statistical Learning

The brain operates as a predictive system, constantly anticipating sensory inputs based on accumulated prior experiences [16]. This neural process relies on resolving differences between expected and actual sensory input through Bayesian inference, known as prediction errors [17]. The internal model is treated as prior probabilities, while sensory inputs serve as likelihoods. By integrating them, the brain generates the most plausible posterior predictions. Crucially, the regulation of prior predictions is modulated by perceptual uncertainty at higher hierarchical levels [18-19], making predictive processing dynamic and adaptive [20].

Music provides a compelling framework to examine predictive processing [21-22], particularly through the lens of statistical learning. Through this process, the brain extracts transitional probabilities from sequences such as music or language, enabling it to predict future stimuli. Neural and computational studies have shown evidence that the brain encodes transitional probabilities of music sequences, modulating neural responses to expected versus unexpected musical events [23-27]. For example, when a musical tone with high transitional probability is predicted, neural responses to the tone are attenuated (for review, see [28]). Remarkably, statistical learning enables even sleeping neonates and infants to identify and segment high-probability sound chunks, such as words or phrases, from speech, despite their lack of prior linguistic knowledge [9-10].

Beyond perception, musical statistical learning plays a critical role in the emergence of embodied emotional engagement. For instance, musical syncopation—a deviation from expected rhythmic patterns—generates prediction errors that elicit motor impulses, such as foot-tapping, as the brain attempts to align internal and external rhythms [29]. This phenomenon, termed music active inference [21-22], highlights the integration of sensory prediction and motor response in reducing perceptual uncertainty in music. Importantly, the complexity of rhythmic patterns influences the brain's capacity for active inference. Overly complex rhythms may inhibit the resolution of prediction errors, while rhythms of intermediate complexity—characterised by a balance of uncertainty and predictability—are more likely to evoke embodied responses [22]. This relationship is often described by an inverted-U shaped curve, where peak pleasure and engagement occur at optimal levels of uncertainty-weighted prediction error. A neuroimaging study reveals that such peak conditions activate reward-related brain regions, including the nucleus accumbens and the orbitofrontal cortex, alongside motor systems such as the basal ganglia and premotor cortex [30]. This principle applies not only to rhythm but also to harmonic progressions. A study demonstrated that specific combinations of uncertainty and surprise—where one is high and the other is low—correspond to peaks in the inverted-U curve of uncertainty-weighted prediction error. These peaks are associated with heightened activity in reward-related brain regions and greater musical enjoyment [31]. These findings illustrate how embodied musical cognition links prediction errors to both motor behavior and dopaminergic reward systems, fostering emotional and aesthetic experiences.

# 3 Three Key insights into Individuality and Creativity

Here, to fully understand how statistical learning contributes to musical creativity and individuality, we propose three key elements: 1) Hierarchical processing: The brain integrates local and global patterns of music sequences, enabling the construction of coherent musical structures, 2) Temporal dynamics: Prediction and uncertainty fluctuate over time, driving adaptive learning and creative exploration, and 3) Embodied Cognition: Musical experiences engage both the brain and the body, with physical responses (e.g., chills, increased heart rate) enriching the aesthetic and emotional dimensions of music. The three components illuminate the interplay between predictive processing, statistical learning, individuality, and creativity in music. They highlight how the brain not only perceives and predicts music but also generates novel compositions and deeply emotional experiences through the dynamic modulation of uncertainty and prediction errors.

### 3.1 Hierarchy

A hierarchical structure is fundamental in language and music [32], reflecting the temporal and spectral organizations of sensory stimuli [33-34]. For example, the hierarchical temporal structure of language rhythm consists of large rhythms around 2Hz corresponding to prosody, under which rhythms around 4-12Hz corresponding to syllables are nested, and further, rhythms around 12-30Hz corresponding to phonemes and onsets of sounds exist [35]. A similar hierarchical structure is present in music rhythms [36], where, for example, at 120 BPM (beats per minute), rhythms corresponding to quarter notes at 2Hz may include rhythms of eighth notes at 4Hz (Fig. 1a, left). Such hierarchical structures can be visualized through the amplitude modulation envelope of sound waveforms (Fig. 1a, right). The hierarchical rhythm structures have been confirmed in various languages such as English [35], Japanese [37-38], Spanish [39], Portuguese [40], French, and German [41], as well as in various musical genres [36]. This suggests that these structures are fundamental characteristics shared by both language and music.

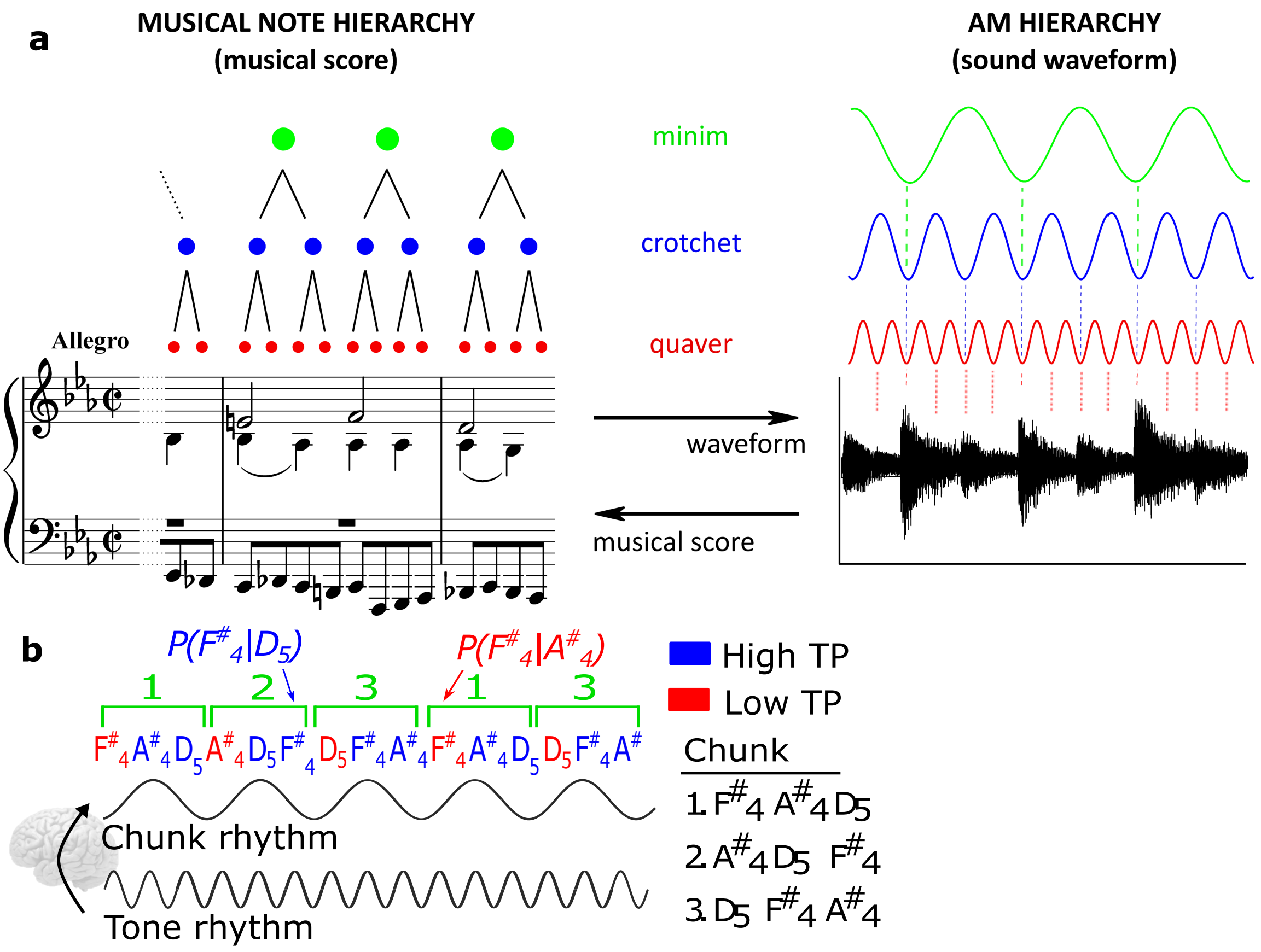

**Fig. 1. Hierarchical structure of musical rhythm (a) and phase synchronization in the brain (b), reprinted from [36, 67].** Neural oscillators phase-synchronize with the hierarchical amplitude modulation inherent in the sound waveform (a). For example, when the duration of a single tone is 333 ms (3 Hz) in (b), the chunk frequency for a sequence of three tones is 1 Hz. When listening to a structured sequence (statistical learning), brain oscillations exhibit phase synchronization not only at the frequency of individual tones but also at the chunk frequency.

Recent studies have demonstrated that the brain extracts and synchronizes with these hierarchical rhythms via phase coupling between neural oscillations and the corresponding frequency bands of auditory input [42-43]. For instance, the brain's delta oscillations phase-synchronize with the prosodic rhythms of speech or the larger rhythms in music, while theta oscillations synchronize with syllabic or finer temporal structures [44]. This synchronization facilitates the decomposition of complex auditory inputs into hierarchical components, supporting both perception and learning.

Interestingly, this process is not solely dependent on AM features of sound waveform. Even in the absence of explicit rhythmic cues in sound waveform, such as accents or strong onsets, the brain generates hierarchical structures by integrating lexical, syntactic, and statistical knowledge [45-46]. For example, during statistical learning, neural oscillations initially synchronize with the frequencies of individual tones or syllables (Fig. 1b). As statistical learning progresses, these oscillations adapt to statistically chunked units, such as words or phrases, reflecting the hierarchical organization of learned patterns [47-48]. The delta rhythm plays a crucial role in auditory learning during infancy. Studies have shown that neural oscillations in infants initially phase-synchronize delta rhythms before transitioning to theta-alpha rhythms, indicating a holistic-to-analytic progression in auditory learning [44]. Infant-directed speech (IDS) and music, characterized by prominent delta rhythms [36], may be more effective than adult-directed speech in facilitating this early learning process [49]. This suggests that hierarchical rhythms in music and IDS may be critical for the initial stages of music and language acquisition.

Hierarchical processing also contributes to creativity, particularly in music [5]. Statistical learning enables the brain to identify and chunk high-probability patterns, which form the basis for generating novel compositions. For example, jazz improvisation frequently incorporates common chord progressions, such as the II–V–I progression, which exhibit high transitional probability and low uncertainty. While these progressions represent shared knowledge at lower hierarchical levels, individual creativity emerges through unique combinations and manipulations of these chunked units (Fig. 1c). A comparative analysis of improvisations by renowned jazz pianists Bill Evans, Herbie Hancock, and McCoy Tyner revealed that lower-hierarchical statistical units were common across all three musicians, whereas higher-hierarchical units were distinct and reflected individual styles [50]. This indicates that while statistical learning provides a foundation for shared musical structures, hierarchical integration fosters individuality in creative expression.

The knowledge and chunk rhythms acquired through statistical learning are also utilized when generating new information. The author conducted simulation experiments using a corpus of child music (i.e., nursery rhymes) from Japan, France, Germany, Korea, and the UK to verify how hierarchical structures of rhythms are acquired through musical statistical learning and how the acquired knowledge is used to generate new music. This was achieved using the HBSL model [15]. This model integrates

Bayesian estimation with Dirichlet prior distribution and Markov processes, allowing the calculation of confidence in probabilities from the inverse of the variance of the prior distribution of transition probabilities. By using normalized values of transition probabilities and confidence, chunking occurs when "confidence * probability" exceeds a constant $\alpha$. The results of the simulation experiments showed that, through 15 trials of statistical learning, the number and hierarchy of chunks increased regardless of language or musical culture. Additionally, it was observed that the probabilistic components of the music generated by the model after learning gradually changed to new ones compared to the original music. Interestingly, it was found that the generated music increasingly contained rhythms in the delta band as learning progressed. This suggests that even without delta band rhythms as acoustic features of the learning information, the brain can chunk information through statistical learning, construct hierarchical rhythm structures internally, and generate information rich in delta band rhythms when generating new information. These results indicate that the statistical learning ability of the brain contributes to the formation of hierarchical structures and new information, which are important for the early development of language and music. This might also explain why nursery rhymes and IDS for children are rich in these rhythms [36].

Furthermore, I generated three HBSL models with different dependencies on sensory signals (hypo: sensory hypo-sensitivity, normal: typical sensitivity, hyper: sensory hyper-sensitivity) to verify how individual difference in sensory sensitivity transforms internal models through statistical learning of a nursery rhyme called “Yuuyake Koyake”. Recently, the understanding of individual developmental characteristics based on the brain's predictive processing has been explained by sensitivity to bottom-up sensory signals from the external environment [51-53]. In this model, the hyper model exhibits fluctuations in the reliability of probabilities (variance of the prior distribution in the internal model) due to stimuli, while the hypo model shows less fluctuation in reliability even with new stimuli. Simulation results showed that hyper-sensitive models, which prioritize bottom-up sensory input, excel at chunking and learning hierarchical rhythms but tend to overfit existing patterns, limiting creative output. Conversely, hypo-sensitive models, less reliant on sensory input, exhibited reduced learning efficiency but generated more novel compositions. These findings underscore a trade-off between learning efficiency and creative generation, highlighting how individual differences in sensory sensitivity influence hierarchical processing and creativity.

Taken together, the hierarchical structures play a crucial role in music as well as language, supporting not only perception and learning but also creativity. Statistical learning contributes to internalizing these structures, enabling the emergence of creativity and individuality. Moreover, individual differences in sensory sensitivity influence the formation and utilization of hierarchical representations, suggesting a dynamic interplay between perceptual processing and creative expression. These insights deepen our understanding of how the brain processes structured auditory input and provide a foundation for exploring developmental and cross-cultural variations in music and language acquisition.

## 3.2 Temporal Dynamics of Predictability

Temporal dynamics of predictability, encompassing fluctuations in uncertainty and prediction error, play a critical role in shaping the creative and emotional dimensions of music [14, 54]. Unlike static predictability, these dynamic variations foster individuality, influencing both the perception and production of music. For instance, when a predictable chord follows an entirely predictable progression, the result may feel monotonous and uninspiring. Conversely, when a predictable chord follows an unpredictable progression, it often elicits a sense of relief or resolution, highlighting the brain's sensitivity to dynamic shifts in predictability.

In a recent study [8], the authors developed the HBSL model that decodes the relationship between perceptual uncertainty (i.e., entropy) and surprise (i.e., prediction error) in musicusing 80,000 chords from Billboard's US pop songs [55]. Using this model, we generated the eight types of chord progressions. Each type is characterized by varying degrees of temporal dynamics of uncertainty (red lines in Fig. 2) and surprise (blue lines in Fig. 2). After listening to each of these eight types of chord progressions, they assessed how they felt about creativity and beauty from the musical chords as well as valence and arousal, using the nine-point Likert scale [56]. The findings indicate that musical chord progressions characterized by the temporal dynamics of high uncertainty and surprise generated heightened feelings of creativity alongside increased arousal. However, the feelings of beauty and valence were more strongly influenced by the temporal dynamics of predictable, low-uncertainty progressions. These results suggest that the feelings of creativity may operate through a somewhat different cognitive mechanism than that of beauty and valence.

Such temporal dynamics evolve over a composer's lifetime, reflecting the progression of creative exploration. For example, an analysis of Beethoven's piano sonatas revealed a gradual increase in musical uncertainty over his lifetime [56]. This progression may suggest that Beethoven's evolving compositional style incorporated increasing levels of unpredictability, potentially as a means of sustaining novelty and complexity in his artwork. The temporal dynamics extend beyond individual lifetimes to influence musical creativity across historical eras. Using the HBSL model, fluctuations in uncertainty and prediction error were analyzed in a large corpus of jazz improvisations spanning 1925 to 2009 [58]. The results demonstrated distinct temporal characteristics in each era, indicating that the individuality of improvisational styles is shaped by epochal changes in musical creativity. These findings suggest that the brain’s capacity for hierarchical statistical learning not only governs individual compositional styles but also underpins broader cultural and temporal shifts in music.

### 3.3 Embodied Cognition

The connection between bodily sensations and emotions can be better understood through the lens of the brain's predictive processing framework. The brain constructs emotions by minimizing the discrepancies between expected signals generated by its internal model and the sensory information received from exteroceptive, interoceptive, and proprioceptive sensations [59]. It has been proposed that interoception, or the perception of internal bodily states such as heart-beat acceleration, emerges from the ongoing refinement of these prediction errors [60-62]. For example, when a musical chord progression concludes abruptly with an unexpected modulation to a different key, this can result in a sudden increase in interoceptive prediction error, such as an accentuated heartbeat. This heightened prediction error may then be addressed by adjusting interoceptive priors, potentially facilitated through deep breathing.

By applying the bodily mapping assessment of emotion [63], the authors developed the new bodily mapping task [64] to investigate emotional responses and the distribution of subjective bodily sensations elicited by eight types of musical chords generated from the HBSL model [8] (Fig. 2). The results demonstrated that musical chords characterized by low uncertainty and high surprise elicited the positive valence, with sensations predominantly localized to the heart region on the bodily map. Furthermore, when a preceding sequence of chords induces prediction errors (i.e., low-probability chord sequences), the resulting pleasure diminished. In contrast, when the preceding sequences were easy to predict, it enhanced feelings of pleasure and beauty, along with interoceptive sensations related to the heart. Specific fluctuations in prediction and uncertainty may facilitate both interoceptive sensations linked to the heart and an enhanced musical emotion. Such embodied cognition may show individual difference depending on mental state and age [65-66].

A recent study also explores the relationship between the experience of creativity and interoceptive bodily sensations, particularly in the heart and stomach regions [66]. The findings highlight the crucial role of heart-related sensations in fostering creative feelings (Fig. 2). This suggests the complex interplay between bodily sensations and feelings of creativity based on predictive processing. Further, using the same chord sequence that controls prediction error and uncertainty, a recent study involving participants aged 15 to over 80 investigated how bodily responses to music evolve over the human lifespan [66]. The findings indicate that heart and abdomen sensations peak around the age of 20, with a gradual decline and diffusion in older age groups. This may underscore the individual difference of embodied cognition of music based on developmental and age-related changes.

In summary, the embodied nature of musical cognition highlights the integral role of the body in shaping individual and creative experiences. By linking prediction errors to embodied cognition, music fosters a deep connection between cognitive processes and physical responses, enabling a richer understanding of creativity and individuality.

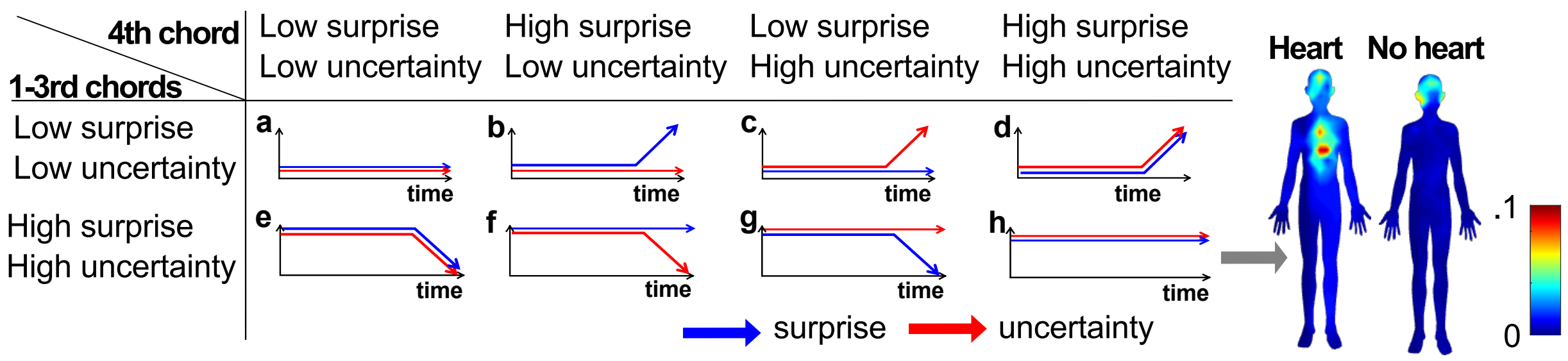

**Fig. 2.** Musical chord progressions with different surprise and uncertainty (left) and body map of the progression **h** in individuals with (Heart) vs. without (No heart) heart sensations (right). Reprinted from Daikoku et al., (2024) [56].

## 4 Summary

This paper sheds light on understanding how individuality and creativity emerge within the framework of the brain's statistical learning. Our findings suggest that musical experience is deeply embedded in predictive processing, where the brain continuously refines its internal model through musical exposure. This dynamic process facilitates not only perception and learning but also the emergence of creative expression and individuality.

One of the key insights is that hierarchical structures in music and language are fundamental to both cognitive organization and creative generation. Notably, our computational modeling demonstrated that hierarchical chunking occurs regardless of linguistic or cultural background, highlighting the universality of this process. The ability to extract and manipulate hierarchical structures allows for the flexible recombination of musical elements, a mechanism that underpins both improvisation and the evolution of musical styles. Moreover, our findings underscore the temporal dynamics of predictability in shaping musical experience. Creativity emerges not merely from static rule-based knowledge but through fluctuations in uncertainty and prediction errors over time. This perspective aligns with previous research indicating that composers progressively explore higher levels of unpredictability throughout their careers, fostering novel and expressive output [55]. The analysis of jazz improvisation and historical shifts in musical composition suggests that statistical learning operates at both individual and collective levels, driving stylistic evolution across time [56]. These insights contribute to a growing understanding of how predictive processing governs not only moment-to-moment musical perception but also long-term creative development. Furthermore, the role of embodied cognition in musical creativity was evident in our findings on interoceptive sensations [66]. These findings support the notion that musical creativity is not purely an abstract cognitive process but one that is grounded in the body's physiological and emotional states.

Taken together, our findings highlight the multifaceted nature of musical creativity, where hierarchical processing, temporal dynamics of predictability, and embodied cognition converge to shape individual and cultural expressions of music. The predictive brain not only perceives and interprets music but also actively generates new musical structures, driven by a balance of prediction error and uncertainty. This study provides a novel framework for understanding creativity as an emergent property of hierarchical statistical learning.

# References

1. Cross, I.: Music, cognition, culture, and evolution. Annals of the New York Academy of sciences 930(1), 28-42 (2001).
2. Leman, M.: Embodied music cognition and mediation technology. MIT press (2007).
3. Putkinen, V., Zhou, X., Gan, X., Yang, L., Becker, B., Sams, M., & Nummenmaa, L.: Bodily maps of musical sensations across cultures. Proceedings of the National Academy of Sciences, 121(5), e2308859121 (2024).
4. Van Der Schyff, D., Schiavio, A., Walton, A., Velardo, V., & Chemero, A. Musical creativity and the embodied mind: Exploring the possibilities of 4E cognition and dynamical systems theory. Music & science, 1, 2059204318792319 (2018).
5. Daikoku, T., Wiggins, G. A., & Nagai, Y.: Statistical properties of musical creativity: Roles of hierarchy and uncertainty in statistical learning. Frontiers in Neuroscience, 15, 640412 (2021).
6. Zioga, I., Harrison, P. M., Pearce, M. T., Bhattacharya, J., & Luft, C. D. B.: From learning to creativity: Identifying the behavioural and neural correlates of learning to predict human judgements of musical creativity. NeuroImage, 206, 116311 (2020).
7. Daikoku, T.: Entropy, uncertainty, and the depth of implicit knowledge on musical creativity: computational study of improvisation in melody and rhythm. Frontiers in Computational Neuroscience, 12, 97 (2018).
8. Daikoku, T., Tanaka, M., & Yamawaki, S.: Bodily maps of uncertainty and surprise in musical chord progression and the underlying emotional response. iScience, 27(4), 109498 (2024).
9. Saffran, J. R., Aslin, R. N., & Newport, E. L.: Statistical learning by 8-month-old infants. Science, 274(5294), 1926-1928 (1996).
10. Teinonen, T., Fellman, V., Näätänen, R., Alku, P., & Huotilainen, M.: Statistical language learning in neonates revealed by event-related brain potentials. BMC neuroscience, 10, 1-8 (2009).
11. Siegelman, N., Bogaerts, L., Christiansen, M. H., & Frost, R.: Towards a theory of individual differences in statistical learning. Philosophical Transactions of the Royal Society B: Biological Sciences, 372(1711), 20160059 (2017).
12. Kidd, E., & Arciuli, J.: Individual differences in statistical learning predict children's comprehension of syntax. Child development, 87(1), 184-193 (2016).
13. Misyak, J. B., & Christiansen, M. H.: Statistical learning and language: An individual differences study. Language Learning, 62(1), 302-331 (2012).

14.Daikoku, T.: Temporal dynamics of statistical learning in children's song contributes to phase entrainment and production of novel information in multiple cultures. Scientific Reports, 13(1), 18041 (2023).

15.Daikoku, T., Kamermans, K., & Minatoya, M.: Exploring cognitive individuality and the underlying creativity in statistical learning and phase entrainment. EXCLI journal, 22, 828 (2023).

16.Friston, K.: The free-energy principle: a unified brain theory?. Nature reviews neuroscience, 11(2), 127-138 (2010).

17.Friston, K.: A theory of cortical responses. Philosophical transactions of the Royal Society B: Biological sciences, 360(1456), 815-836 (2005).

18.Okano, T., Daikoku, T., Ugawa, Y., Kanai, K., & Yumoto, M.: Perceptual uncertainty modulates auditory statistical learning: A magnetoencephalography study. International Journal of Psychophysiology, 168, 65-71 (2021).

19.Daikoku, T., & Yumoto, M.: Order of statistical learning depends on perceptive uncertainty. Current Research in Neurobiology, 4, 100080 (2023).

20.Daikoku, T., & Yumoto, M.: Musical expertise facilitates statistical learning of rhythm and the perceptive uncertainty: A cross-cultural study. Neuropsychologia, 146, 107553 (2020).

21.Koelsch, S., Vuust, P., & Friston, K.: Predictive processes and the peculiar case of music. Trends in cognitive sciences, 23(1), 63-77 (2019).

22.Vuust, P., Heggli, O. A., Friston, K. J., & Kringelbach, M. L.: Music in the brain. Nature Reviews Neuroscience, 23(5), 287-305 (2022).

23.Daikoku, T., Yatomi, Y., & Yumoto, M.: Statistical learning of music-and language-like sequences and tolerance for spectral shifts. Neurobiology of learning and memory, 118, 8-19 (2015).

24.Daikoku, T., Yatomi, Y., & Yumoto, M.: Pitch-class distribution modulates the statistical learning of atonal chord sequences. Brain and cognition, 108, 1-10 (2016).

25.Pearce, M. T., Ruiz, M. H., Kapasi, S., Wiggins, G. A., & Bhattacharya, J.: Unsupervised statistical learning underpins computational, behavioural, and neural manifestations of musical expectation. NeuroImage, 50(1), 302-313 (2010).

26.Pearce, M. T., & Wiggins, G. A.: Auditory expectation: the information dynamics of music perception and cognition. Topics in cognitive science, 4(4), 625-652 (2012).

27.Daikoku, T., Jentschke, S., Tsogli, V., Bergström, K., Lachmann, T., Ahissar, M., & Koelsch, S.: Neural correlates of statistical learning in developmental dyslexia: An electroencephalography study. Biological Psychology, 181, 108592 (2023).

28.Daikoku, T.: Neurophysiological markers of statistical learning in music and language: Hierarchy, entropy and uncertainty. Brain sciences, 8(6), 114 (2018).

29. Witek, M. A., Clarke, E. F., Wallentin, M., Kringelbach, M. L., & Vuust, P.: Syncopation, body-movement and pleasure in groove music. PloS one, 9(4), e94446 (2014).
30. Matthews, T. E., Witek, M. A., Lund, T., Vuust, P., & Penhune, V. B.: The sensation of groove engages motor and reward networks. NeuroImage, 214, 116768 (2020).
31. Cheung, V. K., Harrison, P. M., Meyer, L., Pearce, M. T., Haynes, J. D., & Koelsch, S.: Uncertainty and surprise jointly predict musical pleasure and amygdala, hippocampus, and auditory cortex activity. Current Biology, 29(23), 4084-4092 (2019).
32. Patel, A. D.: Language, music, syntax and the brain. Nature neuroscience, 6(7), 674-681 (2003).
33. Lerdahl, F., & Jackendoff, R.: An overview of hierarchical structure in music. Music Perception, 1(2), 229-252 (1983).
34. Goswami, U.: A neural basis for phonological awareness? An oscillatory temporal-sampling perspective. Current directions in psychological science, 27(1), 56-63 (2018).
35. Leong, V., & Goswami, U.: Acoustic-emergent phonology in the amplitude envelope of child-directed speech. PloS one, 10(12), e0144411 (2015).
36. Daikoku, T., & Goswami, U.: Hierarchical amplitude modulation structures and rhythm patterns: Comparing Western musical genres, song, and nature sounds to Babytalk. Plos one, 17(10), e0275631 (2022).
37. Daikoku, T., Kumagaya, S., Ayaya, S., & Nagai, Y.: Non-autistic persons modulate their speech rhythm while talking to autistic individuals. Plos one, 18(9), e0285591 (2023).
38. Daikoku, T., & Goswami, U.: The Amplitude Modulation Structure of Japanese Infant- and Child-Directed Speech: Longitudinal Data Reveal Universal Acoustic Physical Structures Underpinning Moraic Timing, arXiv preprint arXiv: arXiv preprint arXiv:2503.05645 (2025).
39. Pérez-Navarro, J., Lallier, M., Clark, C., Flanagan, S., & Goswami, U.: Local temporal regularities in child-directed speech in Spanish. Journal of Speech, Language, and Hearing Research, 65(10), 3776-3788 (2022).
40. Araújo, J., Flanagan, S., Castro-Caldas, A., Goswami, U.: The temporal modulation structure of illiterate versus literate adult speech. PLoS One 13, e0205224 (2018).
41. Daikoku, T., Lee, C., & Goswami, U.: Amplitude modulation structure in French and German poetry: universal acoustic physical structures underpin different poetic rhythm structures. Royal Society Open Science, 11(9), 232005 (2024).
42. Gross, J. et al. Speech rhythms and multiplexed oscillatory sensory coding in the human brain. PLOS Biol. 11, e1001752 (2013).

43. Poeppel, D., & Assaneo, M. F.: Speech rhythms and their neural foundations. Nature reviews neuroscience, 21(6), 322-334 (2020).
44. Attaheri, A., et al.: Delta-and theta-band cortical tracking and phase-amplitude coupling to sung speech by infants. NeuroImage, 247, 118698 (2022).
45. Ding, N., Melloni, L., Zhang, H., Tian, X., & Poeppel, D.: Cortical tracking of hierarchical linguistic structures in connected speech. Nature neuroscience, 19(1), 158-164 (2016).
46. Baltzell, L. S., Srinivasan, R., & Richards, V.: Hierarchical organization of melodic sequences is encoded by cortical entrainment. Neuroimage, 200, 490-500 (2019).
47. Smalle, E. H., Daikoku, T., Szmalec, A., Duyck, W., & Mottonen, R.: Unlocking adults' implicit statistical learning by cognitive depletion. Proceedings of the National Academy of Sciences, 119(2), e2026011119 (2022).
48. Ringer, H., Sammler, D., & Daikoku, T.: Neural tracking of auditory statistical regularities is reduced in adults with dyslexia. Cerabral Cortex, 35(2), bhaf042 (2025).
49. Leong, V., Kalashnikova, M., Burnham, D., & Goswami, U.: The temporal modulation structure of infant-directed speech. Open Mind, 1(2), 78-90 (2017).
50. Daikoku, T.: Musical creativity and depth of implicit knowledge: spectral and temporal individualities in improvisation. Frontiers in computational neuroscience, 12, 89 (2018).
51. Pellicano, E., and Burr, D.: When the world becomes 'too real': a Bayesian explanation of autistic perception. Trends Cogn. Sci. 16, 504-510 (2012).
52. Lawson, R. P., Rees, G., and Friston, K. J.: An aberrant precision account of autism. Front. Hum. Neurosci. 8:302 (2014).
53. Philippsen, A., Tsuji, S., & Nagai, Y.: Simulating developmental and individual differences of drawing behavior in children using a predictive coding model. Frontiers in Neurorobotics, 16, 856184 (2022).
54. Abdallah, S., & Plumbley, M.: Information dynamics: patterns of expectation and surprise in the perception of music. Connection Science, 21(2-3), 89-117 (2009).
55. Burgoyne, J. A., Wild, J., & Fujinaga, I.: An Expert Ground Truth Set for Audio Chord Recognition and Music Analysis. In ISMIR, 11, 633-638 (2011).
56. Daikoku, T., & Tanaka, M.: Body Maps for Feeling of Creativity in Musical Chord Progression. arXiv preprint arXiv:2410.11885 (2024).
57. Daikoku, T.: Depth and the uncertainty of statistical knowledge on musical creativity fluctuate over a composer's lifetime. Frontiers in computational neuroscience, 13, 27 (2019).
58. Daikoku, T.: Temporal dynamics of uncertainty and prediction error in musical improvisation across different periods. Scientific Reports, 14(1), 22297 (2024).

59.Seth, A. K.: Interoceptive inference, emotion, and the embodied self. Trends in cognitive sciences, 17(11), 565-573 (2013).

60.Barrett, L. F., & Simmons, W. K.: Interoceptive predictions in the brain. Nature reviews neuroscience, 16(7), 419-429 (2015).

61.Khalsa, S. S., Craske, M. G., Li, W., Vangala, S., Strober, M., & Feusner, J. D.: Altered interoceptive awareness in anorexia nervosa: Effects of meal anticipation, consumption and bodily arousal. International Journal of Eating Disorders, 48(7), 889-897 (2015).

62.Ainley, V., Apps, M. A., Fotopoulou, A., & Tsakiris, M.: 'Bodily precision': a predictive coding account of individual differences in interoceptive accuracy. Philosophical Transactions of the Royal Society B: Biological Sciences, 371(1708), 20160003 (2016).

63.Nummenmaa, L., Glerean, E., Hari, R., & Hietanen, J. K.: Bodily maps of emotions. Proceedings of the National Academy of Sciences, 111(2), 646-651 (2014).

64.Daikoku, T., & Tanaka, M.: Protocol to visualize a bodily map of musical uncertainty and prediction. STAR protocols, 5(4), 103473 (2024).

65.Tanaka, M., & Daikoku, T.: Music-related Bodily Sensation Map in Individuals with Depressive Tendencies. bioRxiv, 2024.06. 14.599050 (2024). doi.org/10.1101/2024.06.14.599050

66.Tanaka, M., and Daikoku, T.: Lifetime changes in Body map based on music prediction, b bioRxiv 2025.02.02.636100, (2025). doi: https://doi.org/10.1101/2025.02.02.636100.

67.大黒達也. (2023). 音系列のリズムと構造の予測・認知・生成. 日本音響学会誌, 80(1), 18-24.